\documentclass{article}

\usepackage[final]{neurips}

\usepackage[utf8]{inputenc} 
\usepackage[T1]{fontenc}    
\usepackage{hyperref}       
\usepackage{url}            
\usepackage{booktabs}       
\usepackage{amsfonts}       
\usepackage{nicefrac}       
\usepackage{microtype}      
\usepackage{lipsum}
\usepackage{amsmath}
\usepackage{graphicx}
\usepackage{tikz}
\usepackage{enumitem}
\usetikzlibrary{positioning,arrows.meta,calc,fit}
\usepackage{float}
\usepackage{caption}
\usepackage{tcolorbox}
\usepackage[table]{xcolor}

\setcitestyle{authoryear,round}

\usepackage{xcolor}

\definecolor{citepurple}{RGB}{0,0,128}

\hypersetup{
    colorlinks=true,
    linkcolor=black,
    citecolor=citepurple,
    urlcolor=black
}

\graphicspath{ {./images/} }

\title{NIM4-ASR: Towards Efficient, Robust, and Customizable Real-Time LLM-Based ASR}

\makeatletter
\renewcommand{\@noticestring}{$\textsuperscript{*}$Equal Contribution.} 
\makeatother
\author{
  Yuan Xie$^{*}$, \And 
  Jiaqi Song$^{*}$, \And 
  Guang Qiu, \And 
  Xianliang Wang, \And
  Kai Qiao, \And 
  Junfeng Yuan \And \\[0.20em]
  Shengqing Liu, \And 
  Yi Zhang, \And
  Bowen Chen, \And
  Ming Lei, \And 
  Jie Gao, \And 
  Jie Wu \\[0.75em]
  \textnormal{Advanced Intelligent Systems Group, NIO}  \\[0.75em]
  \texttt{\{ryan.xie2, jiaqi.song2\}@nio.com}
}

\begin{document}
\maketitle
\thispagestyle{empty}

\definecolor{linkblue}{RGB}{0,51,102}
\vspace{-3.0em}
\begin{center}
\textbf{Project page: }
\href{https://yuanx9.github.io/NIM4-ASR/}
{\textcolor{linkblue}{\texttt{https://yuanx9.github.io/NIM4-ASR/}}}
\end{center}
\vspace{0.5em}

\begin{abstract}
 
Integrating large language models (LLMs) into automatic speech recognition (ASR) has become a mainstream paradigm in recent years. Although existing LLM-based ASR models demonstrate impressive performance on public benchmarks, their training remains predominantly data-driven, leaving key practical challenges insufficiently addressed---particularly limited downward scalability in resource-constrained deployments and hallucinations under acoustically challenging conditions. 
To address these issues, we present \textbf{NIM4-ASR}, a production-oriented LLM-based ASR framework optimized for both efficiency and robustness. Grounded in a principled delineation of functional roles between the encoder and the LLM, we redesign the multi-stage training paradigm to align each module with its intended capability boundary. Specifically, we reformulate the pre-training architecture and objective to mitigate the modality gap and improve parameter efficiency; introduce an iterative asynchronous SFT stage to preserve acoustic fidelity and constrain representation drift; and design an ASR-specialized reinforcement learning stage to further enhance recognition quality and robustness. We additionally incorporate a suite of production-oriented optimizations, including robustness under noisy and silent conditions, real-time streaming inference, and hotword customization via retrieval-augmented generation (RAG). 
Experiments show that NIM4-ASR achieves state-of-the-art performance on multiple public benchmarks with merely 2.3B parameters, while substantially outperforming larger-scale competitors on internal benchmarks---particularly in entity-intensive real-world scenarios. NIM4-ASR further supports million-scale hotword customization via RAG with sub-millisecond retrieval latency, enabling efficient adaptation to emerging entities and personalized user requirements.

\end{abstract}


\section{Introduction}
\label{sec-Introduction}
With the rapid advancement of large language models (LLMs), the prevailing paradigm of automatic speech recognition (ASR) is undergoing a transition from classical architectures~\citep{graves2006connectionist,chorowski2015attention,chan2016listen,graves2012sequence} to the encoder--adaptor--LLM framework~\citep{bai2024seed,an2025funaudio}. Over the past two years, a series of LLM-based ASR models, including Seed-ASR~\citep{bai2024seed}, Fun-ASR~\citep{an2025funaudio}, FireRedASR series~\citep{xu2025fireredasr,xu2026fireredasr2}, Voxtral~\citep{liu2025voxtral}, Index-ASR~\citep{song2025index}, and Qwen3-ASR~\citep{shi2026qwen3}, have achieved promising performance on public ASR benchmarks.

Compared with classical ASR models that are primarily optimized for acoustic-to-lexical transduction, LLM-based ASR benefits from the rich linguistic priors and contextual modeling capacity inherited from large-scale language model pretraining~\citep{fathullah2024prompting}. The LLM's strong language modeling capacity and contextual coherence modeling help resolve acoustic and lexical ambiguities, yielding transcriptions that are more fluent and semantically coherent. Furthermore, LLMs encode extensive world knowledge during large-scale pretraining, substantially improving the recognition of rare named entities, technical terminology, and domain-specific expressions that classical ASR models frequently misrecognize~\citep{wang2025contextasr}. 
Overall, incorporating LLMs helps bridge acoustic modeling with semantic understanding~\citep{an2025funaudio,hono2024integrating}, leading to enhanced robustness to acoustically ambiguous inputs such as noise and accent variations, as well as improved cross-domain generalization.
Despite these advantages, LLM-based ASR still faces several key limitations in real-world scenarios.

\textbf{1. Limited downward scalability.} In deployment, particularly for real-time speech interfaces, lightweight ASR models are favored for their lower inference latency and computational cost. However, the downward scalability of LLM-based ASR appears disappointing: lightweight variants such as Qwen3-ASR-0.6B and Fun-ASR-nano exhibit substantial performance gaps relative to their full-scale counterparts. Beyond the degradation ordinarily expected from model downscaling, LLM-based ASR models carry an additional structural cost from the modality tax~\citep{aghajanyan2023scaling,zhang2026instruction}: a non-trivial number of parameters are devoted to cross-modal alignment rather than the ASR task itself. This overhead leaves lightweight LLMs with less effective capacity, imposing a disproportionate performance degradation~\citep{endo2025downscaling}.

\textbf{2. Hallucination.} Beyond the intrinsic hallucination tendencies of autoregressive LLMs, the encoder--adaptor--LLM joint-training paradigm introduces additional risks~\citep{bai2024hallucination,zhou2024mitigating,xie2026rethinking}. During joint optimization, the encoder is progressively pulled toward the LLM's optimization objective under the influence of its stronger gradients and linguistic priors, causing its representations to gradually shift toward the LLM's text feature space (i.e., representation drift). As a result, the encoder may increasingly rely on linguistic shortcuts at the expense of fine-grained acoustic fidelity, exacerbating hallucinations under acoustically ambiguous conditions~\citep{park2025evaluating}. In task-oriented in-vehicle speech interaction scenarios, hallucinations can cascade through the downstream pipeline and trigger unintended actions~\citep{tay2026back}.

\textbf{3. Lack of production-ready hotword customization.} Existing LLM-based ASR systems lack mature hotword customization solutions comparable to N-gram language model rescoring methods~\citep{song2019l2rs,kuo2022correcting} widely adopted in classical ASR systems. Such customization support is indispensable for accurately transcribing personalized entities with similar pronunciations~\citep{an2025funaudio,lei2025contextualization}, including homophonous location names, media titles, and emerging proper nouns that often reside in the long tail of LLMs' pretraining distribution.

To address the aforementioned limitations, we propose \textbf{NIM4-ASR (NOMI Intelligence Model 4.0-ASR)}, a production-oriented LLM-based ASR framework optimized for both efficiency and robustness. 
NIM4-ASR adopts a redesigned multi-stage training paradigm that reduces the modality gap between speech and text while explicitly delineating the functional roles of the encoder and the LLM~\citep{xie2026rethinking}. Specifically, we redesign a module-aware pre-training scheme that aligns training objectives with the intrinsic characteristics of each component. This encourages the encoder to produce low-entropy, peaky representations that narrow the modality gap, reducing the LLM capacity required for cross-modal alignment and improving parameter efficiency. We then develop an Iterative Asynchronous SFT (IA-SFT) stage between alignment and joint SFT, which strengthens cross-modal alignment while preserving functional decoupling across modules, thereby mitigating representation drift and suppressing hallucinations. Additionally, we incorporate an ASR-specialized reinforcement learning (RL) strategy to further enhance recognition quality and robustness.
Beyond the training-side design, NIM4-ASR also incorporates a series of production-oriented enhancements for practical deployment, including robustness under noisy and silent conditions, real-time streaming inference, and scalable hotword customization via retrieval-augmented generation (RAG). Finally, we conduct extensive evaluations on diverse Mandarin and English benchmarks, demonstrating that NIM4-ASR achieves state-of-the-art (SOTA) performance on several benchmarks with only 2.3B parameters. Our key contributions are summarized as follows:

\begin{itemize}[leftmargin=*]
\item \textbf{Principled multi-stage training paradigm.} We propose a principled multi-stage training paradigm that reduces the modality gap and preserves module-specific functional specialization for improved efficiency and robustness. We further introduce an ASR-specialized RL stage, which brings additional gains in recognition accuracy and hallucination mitigation.

\item \textbf{Optimized streaming support.} 
We cultivate the encoder's native streaming capability from pre-training and introduce a decoupled streaming inference strategy that separates encoder and LLM execution, enabling both responsive real-time transcription and stable final outputs.


\item \textbf{Phoneme-level RAG for hotword customization.} Building on Fun-ASR, we improve the phoneme-level retrieval algorithm with an emphasis on retrieval precision and latency, enabling million-scale hotword customization with sub-millisecond retrieval latency while preserving high retrieval precision.

\item \textbf{Comprehensive evaluation.} We conduct comprehensive evaluations across 25 benchmarks (15 public and 10 internal), showing that NIM4-ASR can achieve SOTA performance on multiple benchmarks with only 2.3B parameters, validating its parameter efficiency and strong robustness.

\end{itemize}

\section{Methodology}
\label{sec-Methodology}

\subsection{Architecture}
\label{sec-Architecture}

\begin{figure}[htbp]
    \centering
    \includegraphics[width=0.9\linewidth]{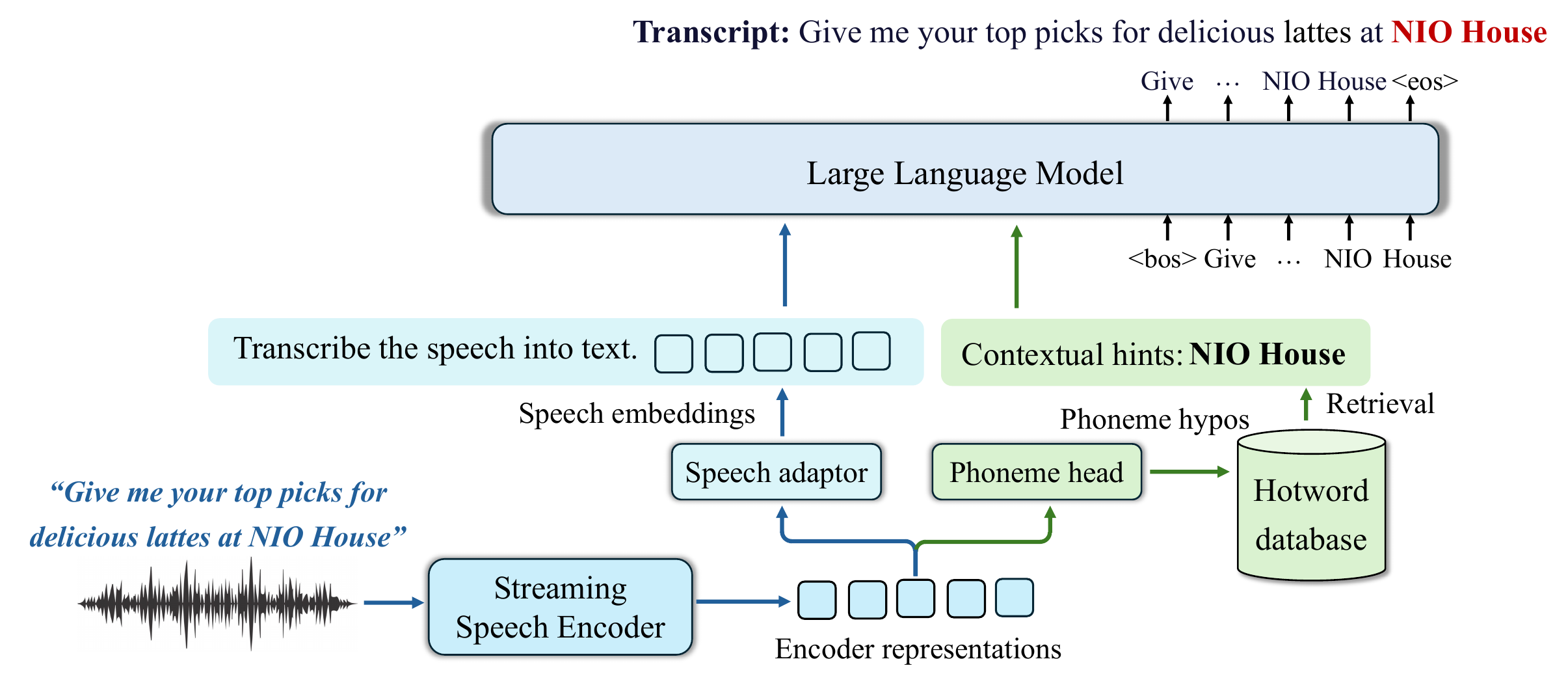}
    \caption{The overall architecture of NIM4-ASR.}
    \label{fig1}
\end{figure}

As shown in Figure~\ref{fig1}, NIM4-ASR follows a modular encoder–adaptor–LLM architecture comprising four components. Before being fed into the model, raw speech is first converted into 80-dimensional log-Mel spectrograms using a 25\,ms window with a 10\,ms frame shift, followed by global mean and variance normalization. The details of the four main components are described below:

\begin{itemize}[leftmargin=*,itemsep=2pt]

\item \textbf{Streaming speech encoder.}
Our encoder adopts the Conformer encoder architecture from FireRedASR-AED~\citep{xu2025fireredasr}, consisting of a 4x downsampling convolutional module followed by a stack of Conformer blocks~\citep{gulati2020conformer}, with approximately 600~M parameters in total. The encoder converts acoustic features into continuous representations at a frame rate of 25~Hz (40~ms temporal resolution). 
To support low-latency online decoding, we convert it into a chunk-based streaming encoder by simulating streaming constraints during training. 

\item \textbf{Speech adaptor.}
The speech adaptor consists of a two-layer MLP that maps the encoder representations into the LLM’s input embedding space. Before projection, we apply a 4x downsampling by concatenating four consecutive frames along the feature dimension to shorten the sequence length. After downsampling, the frame rate is reduced to 6.25~Hz, corresponding to a temporal resolution of 160~ms per token.

\item \textbf{Phoneme-level CTC head and RAG module.}
The phoneme-level CTC head (hereafter referred to as the CTC head or phoneme head) serves as the acoustic front-end of the RAG module, comprising a three-layer MLP. It decodes encoder representations into phoneme hypotheses via greedy decoding. Based on these hypotheses, our retrieval algorithm searches the hotword database to retrieve matching entries, which are then injected into the prompt as contextual hints for the LLM. Further details of the RAG module are provided in Section~\ref{sec-hotword_RAG}.

\item \textbf{LLM decoder.}
The decoder is initialized from Qwen3-1.7B~\citep{yang2025qwen3} and generates the final transcription conditioned on both speech embeddings and optional retrieved hotword hints.

\end{itemize}

\subsection{Training Recipe}
\label{sec-Training_recipe}

\begin{figure}[htbp]
    \centering
    \includegraphics[width=1.0\linewidth]{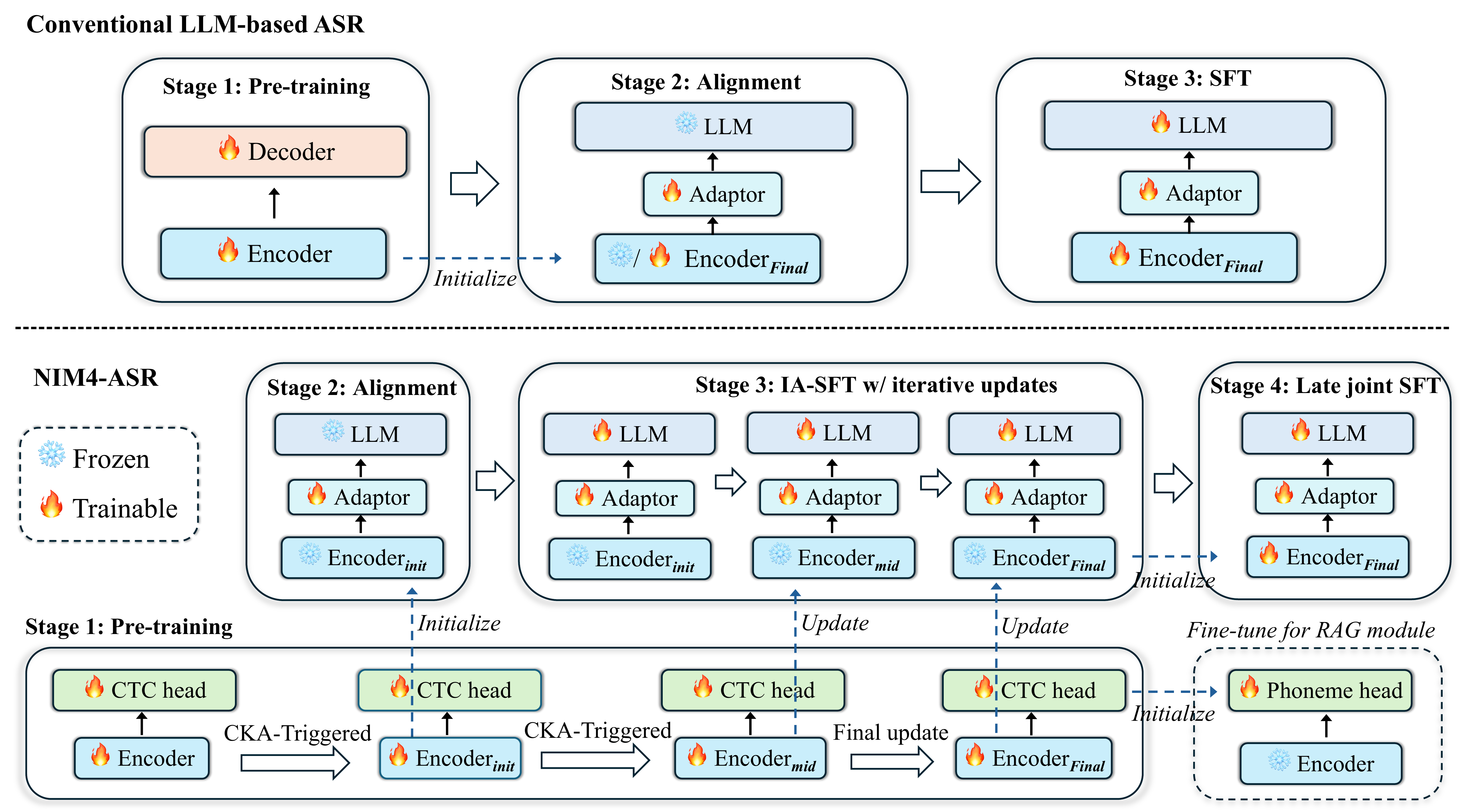}
    \caption{Comparison of training pipelines from encoder pretraining to joint SFT for conventional LLM-based ASR and our NIM4-ASR.}
    \label{fig2}
\end{figure}

In contrast to most prior work driven primarily by empirical fine-tuning, we begin with a principled analysis of the practical limitations of current LLM-based ASR systems and their underlying causes~\citep{xie2026rethinking}, revealing that the cross-modal gap and representation drift remain insufficiently addressed. Based on these insights, we comprehensively redesign the training pipeline. As illustrated in Figure~\ref{fig2}, the methodological advances of NIM4-ASR center on four core training stages: encoder pretraining, alignment, IA-SFT, and late joint SFT. Beyond this four-stage pipeline, context SFT and RL are further incorporated after late joint SFT to strengthen contextual modeling and robustness. The detailed procedures are described below.

\subsubsection{Stage 1: Encoder Pre-training}
\label{sec-Stage1}
To reduce the modality gap between encoder representations and the LLM embedding space, we adopt an improved variant of Connectionist Temporal Classification (CTC)~\citep{graves2006connectionist}---namely CR-CTC~\citep{yao2024cr}---as the pretraining objective. As illustrated in Figure~\ref{fig2}, the model architecture during pretraining consists of the encoder paired with a CTC head. In contrast to the Attention-based Encoder-Decoder (AED) commonly used in prior work~\citep{xu2025fireredasr,xu2026fireredasr2}, CTC encourages the encoder to produce low-entropy, phoneme-discriminative representations that align more naturally with the LLM's embedding space, thereby reducing cross-modal alignment overhead and reserving more model capacity for the ASR task.

Furthermore, we shift the supervision labels from character level to phoneme level~\citep{yusuyin2025whistle}, explicitly dedicating the encoder's capacity to acoustic-to-phoneme mapping rather than premature semantic anchoring, while encouraging the LLM to focus more on semantic reasoning. This design achieves a cleaner decoupling of acoustic modeling from semantic reasoning, improving role specialization of both modules. Moreover, adopting phoneme prediction as the pretraining objective encourages the encoder to learn low-level acoustic representations with weak language dependence, offering greater potential for extending to new languages and dialects.

To endow the encoder with native streaming capability, we incorporate the dynamic-chunk mechanism during pretraining~\citep{zhang2020unified}. Specifically, the encoder processes full utterances under chunk-wise streaming constraints, where the chunk size and the number of visible left-context chunks are dynamically sampled for each batch. This exposes the encoder to a wide range of streaming configurations, enabling flexible operation that accommodates varying latency budgets across different deployment scenarios.

\subsubsection{Stage 2: Alignment \& Stage 3: IA-SFT}
\label{sec-Stage2-3}
In conventional training paradigms, alignment and joint SFT are performed sequentially after pretraining fully completes. As shown in Figure~\ref{fig2}, we propose an encoder iteration mechanism for NIM4-ASR that allows alignment to begin before pretraining completes, while IA-SFT is launched upon alignment completion and proceeds asynchronously alongside the remaining pretraining process. To decide when to initialize or update the encoder used by alignment and IA-SFT, we track encoder representation dynamics using Centered Kernel Alignment (CKA)~\citep{kornblith2019similarity}, which compares the evolving encoder against the reference checkpoint that is initialized and periodically updated throughout pretraining. Given two sets of encoder representations $E^{(a)}, E^{(b)}$ extracted from the same evaluation set, CKA is defined as
\begin{equation}
\text{CKA}(E^{(a)}, E^{(b)}) = \frac{ \langle \tilde{K}^{(a)}, \tilde{K}^{(b)} \rangle_F }{ \sqrt{ \langle \tilde{K}^{(a)}, \tilde{K}^{(a)} \rangle_F \cdot \langle \tilde{K}^{(b)}, \tilde{K}^{(b)} \rangle_F } },
\label{eq:cka}
\end{equation}
where $\tilde{K}^{(a)}$ and $\tilde{K}^{(b)}$ are centered Gram matrices calculated via $\tilde{K}^{(x)} = CE^{(x)} E^{(x)\top}C$. The centering matrix is defined as $C = I_L - \frac{1}{L}J_L$, where $I_L$ is the identity matrix and $J_L$ is the all-ones matrix. CKA measures the geometric consistency of representation spaces, invariant to orthogonal transformation and isotropic scaling.

\noindent\textbf{Stage 2: Alignment.} We start monitoring the encoder after pretraining reaches 500k steps, at which point the encoder begins to exhibit a relatively stable optimization trend. The encoder at 500k steps is snapshotted as the initial reference checkpoint, and CKA is evaluated every 10k pretraining steps thereafter. When the CKA score between the evolving encoder and the current reference checkpoint first falls below the predefined threshold\footnote{In this work, we empirically set this threshold to 0.975 based on global CKA dynamics during pretraining, balancing meaningful representation changes against disruptive representation shifts.}, we snapshot the corresponding encoder to initialize alignment and simultaneously update the reference checkpoint. During alignment, both the encoder and LLM are frozen, and only the adaptor is trained. In our setup, this first trigger occurs at approximately 1.01M pretraining steps, and the alignment stage runs for 1.3M steps.

\noindent\textbf{Stage 3: IA-SFT.} After alignment completes, we perform IA-SFT as an intermediate stage before joint SFT. IA-SFT keeps the encoder frozen and trains the adaptor--LLM stack across a sequence of encoder snapshots produced by the asynchronous pretraining process. The procedure is as follows:

\begin{itemize}[leftmargin=*]
\item \textbf{(i) Initialization \& monitoring.} IA-SFT begins after alignment completes, training for 1M steps with the encoder inherited from alignment, while encoder pretraining continues in parallel. The CKA evaluation resumes from the previously updated reference checkpoint and continues every 10k pretraining steps, monitoring the representation shift.

\item \textbf{(ii) CKA-triggered update.} Whenever the CKA score drops below the predefined threshold, the snapshot of the current pretraining encoder is hot-swapped into the IA-SFT branch, and the reference checkpoint is updated accordingly.

\item \textbf{(iii) Final update.} The update cycle (ii) repeats until pretraining reaches its 2M-step maximum. When pretraining completes, a final encoder update is applied regardless of the CKA score, and IA-SFT runs for the final 2M steps.

\end{itemize}

In our implementation, IA-SFT trains for 1M steps using the encoder checkpoint at 1.01M pretraining steps, another 1M steps using the encoder checkpoint at 1.32M pretraining steps, and a final 2M steps using the fully pretrained encoder---totaling 4M steps across three encoder versions. During IA-SFT, the encoder remains frozen but is periodically updated from the asynchronous pretraining process, thus maintaining acoustic grounding. This allows the model to deepen cross-modal alignment without the risk of representation drift. From a curriculum learning perspective, IA-SFT progressively exposes the LLM to refined encoder representations, allowing it to learn invariant patterns and achieve greater robustness to acoustic perturbations. Furthermore, since alignment and IA-SFT run asynchronously alongside pretraining, the overall training pipeline remains time-efficient.

\subsubsection{Stage 4: Late Joint SFT}
\label{sec-Stage4}

After the completion of both encoder pretraining and IA-SFT, a robust initial cross-modal mapping between speech representations and the LLM embedding space has been established. We then perform late joint SFT, in which the encoder, adaptor, and LLM are jointly optimized in an end-to-end manner. Compared with conventional joint training, the risk of representation drift induced by LLM gradients is substantially reduced, as the preceding stages have already minimized the modality gap. Consequently, these gradients serve primarily as fine-tuning signals that seamlessly refine acoustic-to-phoneme mapping and phoneme-to-semantic grounding. 
From a geometric perspective, the preceding alignment stages have established a stable cross-modal manifold, placing subsequent optimization in a low-curvature region of the loss landscape. Within this regime, gradient updates act as local refinements to decision boundaries and manifold geometry rather than inducing large-scale topological restructuring.

Following late joint SFT, all subsequent training stages, including context SFT and RL, are conducted in a fully end-to-end manner. With modality alignment concerns largely resolved in prior stages, the model can devote its full capacity to refining complex cross-modal reasoning and long-context interaction, progressively deepening the integration of acoustic perception and semantic modeling.

\subsubsection{Stage 5: Context SFT}
\label{sec-Stage5}

Following joint SFT, we introduce a context SFT stage~\citep{bai2024seed,an2025funaudio,song2025index} to strengthen the model's ability to leverage contextual information---a capability essential for hotword customization in LLM-based ASR systems. In this stage, we first construct a keyword set $S$ from the training corpus. All transcripts are parsed to extract candidate phrases, which are then filtered by Qwen3-30B-A3B-Instruct~\citep{yang2025qwen3} to retain named entities such as personal names, POI (points of interest), media names and proper nouns.
During training, we increase the sampling ratio of long-duration utterances and probabilistically inject keywords sampled from $S$ into the prompt as contextual hints, following the template below:



\begin{tcolorbox}[colback=gray!10, colframe=black, title=NIM4-ASR Prompt Template]
\small
{\fontfamily{zi4}\selectfont

\textbf{<|im\_start|>}system\\
You are a Multilingual Speech Recognition Model.\textbf{<|im\_end|>}\\[2pt]

\textbf{<|im\_start|>}user\\
Transcribe the speech into text <speech>.\\
Contextual hints: \{context\}.\textbf{<|im\_end|>}\\[2pt]

\textbf{<|im\_start|>}assistant\\
<transcription>\textbf{<|im\_end|>}

}
\end{tcolorbox}

For each training instance, we first retrieve relevant keywords from $S$ present in the transcript. Additionally, for each keyword, we retrieve another keyword from $S$ with identical or highly similar pronunciation to serve as a distractor context with a certain probability. Both relevant keywords and distractors are concatenated and then added to the \texttt{\{context\}} field. The inclusion of distractors discourages the LLM from over-relying on contextual cues at the expense of semantic plausibility. During this stage, the encoder, adaptor and the LLM are jointly trained.


It is worth noting that our context SFT focuses on phrase-level contextual cues~\citep{an2025funaudio,shi2026qwen3} rather than the sentence- or dialogue-level context~\citep{bai2024seed}, as this stage is designed specifically for hotword customization rather than cross-turn dialogue consistency. For multi-turn scenarios, keywords extracted from dialogue history can also be appended to the current prompt. This strategy preserves critical contextual information in a compact form, while maintaining lower inference latency than sentence-level alternatives.

\subsubsection{Stage 6: ASR Specialized RL}
\label{sec-Stage6}

To further improve transcription quality, we introduce an ASR-specialized RL stage based on Group Relative Policy Optimization (GRPO)~\citep{shao2024deepseekmath} that directly optimizes sequence-level transcription behavior using verifiable rewards. In contrast to supervised objectives that rely on token-level teacher-forcing, RL evaluates complete hypotheses and directly optimizes sequence-level transcription behavior, improving recognition accuracy, hallucination robustness, and context-sensitive keyword recognition.

Given an input audio $q$ with ground truth $y$, the policy model independently samples a group of $K$ candidate hypotheses $\{\tau_1, \ldots, \tau_K\} \sim \pi_\theta(\cdot \mid q)$, while each hypothesis is evaluated using a set of ASR-specific reward functions. 

\begin{itemize}[leftmargin=*]
\item \textbf{Accuracy reward:} We apply a unified text normalization pipeline to both the generated hypotheses and the ground truth, and then compute the character error rate (CER) of each hypothesis. The reward function is defined as $R_{\text{acc}}(\tau,y) = \exp(-\alpha \cdot \mathrm{CER(\tau,y)})$, where $\alpha$ is set to $2.0$ in our experiments. This reward is bounded within $(0, 1]$, and its exponential form amplifies the differences among low-CER hypotheses, encouraging fine-grained optimization on near-correct transcriptions. For high-CER regions (e.g., $\mathrm{CER} > 1.0$), the function requires no clipping and still preserves monotonic reward ordering, which is essential for computing within-group advantages in GRPO.

\item \textbf{Hallucination reward:} We apply mixed-granularity tokenization (character-level for Chinese, word-level for English) to both hypothesis and ground truth, then compute their lengths. The hallucination reward $R_{\text{hallu}}(\tau,y)=-1$ if the hypothesis length exceeds 2× or falls below 0.5× the ground truth length; otherwise $R_{\text{hallu}}(\tau,y)=0$.

\item \textbf{Context reward:} For each training sample, we use Qwen3-30B-A3B-Instruct to annotate 0--2 entity keywords per sample. During training, each sample randomly selects a subset of keywords and injects them into the prompt as contextual hints. For each selected keyword, we check whether it appears in the hypothesis: a hit yields a reward of $+0.5$, while a miss incurs a penalty of $-0.5$. The cumulative score across all valid contexts is then averaged to obtain the sample-level context reward $R_{\text{context}}(\tau,y)$. Notably, we also define a list of important keywords. Whenever a sample contains any important keyword, that keyword is included in the reward computation, regardless of whether it is provided in the prompt.

\end{itemize}

Finally, the total reward is given by: 
\begin{equation}
R(\tau,y)=R_{\text{acc}}(\tau,y) + 0.5R_{\text{hallu}}(\tau,y) + 0.5R_{\text{context}}(\tau,y). 
\end{equation}

\paragraph{Reinforcement learning algorithm.} 

Following GRPO, we compute the group-normalized advantage
\begin{equation}
\hat{A}_{i,t}
=
\frac{
R(\tau_i,y) - \mathrm{mean}(\{R(\tau_j,y)\}_{j=1}^{K})
}{
\mathrm{std}(\{R(\tau_j,y)\}_{j=1}^{K}) + \epsilon
}.
\end{equation}
where $\epsilon$ is a small constant for numerical stability.
Denote $\theta_{\mathrm{old}}$ as the policy parameters at the beginning of each optimization step, $\varepsilon$ as the clipping range, and $\beta$ as the KL penalty coefficient. The GRPO objective is defined as
\begin{equation}
\label{eq:grpo}
\mathcal{J}_{\mathrm{GRPO}}
=
\frac{1}{K}
\sum_{i=1}^{K}
\frac{1}{|\tau_i|}
\sum_{t=1}^{|\tau_i|}
\min \Big(
    r_{i,t}(\theta)\,\hat{A}_{i,t},
    \;
    \mathrm{clip}\!\big(r_{i,t}(\theta),\, 1-\varepsilon,\, 1+\varepsilon\big)\,\hat{A}_{i,t}
\Big)
-
\beta \, D_{\mathrm{KL}}(\pi_\theta \| \pi_{\mathrm{ref}}),
\end{equation}
where
\begin{equation}
r_{i,t}(\theta)
=
\frac{
\pi_\theta(\tau_{i,t} \mid q, \tau_{i,<t})
}{
\pi_{\theta_{\mathrm{old}}}(\tau_{i,t} \mid q, \tau_{i,<t})
}.
\end{equation}

\paragraph{RL training framework.}
We implement an RL training pipeline tailored for LLM-based ASR. For each training batch, the policy model encodes input utterances into speech embeddings, which are reused across both rollout generation and policy model log-probability computation to avoid redundant computation. 
During policy rollout, we leverage vLLM~\citep{kwon2023efficient} to efficiently sample $K$ hypotheses conditioned on the speech embeddings and the instruction prompt. The sampled hypotheses are then scored by the reward functions described above. Policy optimization is conducted in a DeepSpeed ZeRO~\citep{rajbhandari2020zero} distributed training setup, where token-level log-probabilities are computed under both the policy model and the reference model. The reference model remains frozen throughout training, providing a stable anchor for KL regularization and preventing excessive policy drift. After each optimization step, the updated policy weights are synchronized to the vLLM rollout engine, ensuring that hypothesis sampling remains on-policy.

Considering that ASR models typically employ deterministic decoding at inference time, we adopt a cosine-annealed temperature schedule for rollout sampling, gradually decaying from 1.0 to 0.7. In the early stages of RL training, the high temperature encourages diverse hypothesis generation, allowing the reward signal to explore a broader range of transcription behaviors. As training progresses, the temperature is smoothly reduced, progressively reinforcing top-1 path quality to ensure strong performance under deterministic decoding at inference.

\subsubsection{Additional Stage: Phoneme Head Training for RAG}
\label{sec-bonus-stage}

After completing the RL stage, the main training pipeline is concluded. We then introduce an additional stage to train the phoneme head required by the RAG module illustrated in Figure~\ref{fig1}. In this stage, the encoder inherits its structure and weights from the post-RL checkpoint and remains frozen, while the phoneme head is initialized from the pretrained CTC head and remains trainable (see Figure~\ref{fig2}). The training objective and configuration are consistent with those used in pretraining. After fine-tuning, the phoneme head can convert encoder representations into phoneme hypotheses for the subsequent retrieval module.

\subsection{Training Setup}
This section presents additional implementation details, including training tricks and settings.

\textbf{Robustness enhancement under noisy and silent conditions.} In the first five training stages, we apply several data augmentation tricks to improve model robustness. In addition to standard SpecAugmentation~\citep{park2019specaugment} and speed perturbation, we randomly inject realistic acoustic disturbances, such as babble noise, vehicle noise, and background music, into 20\% of clean training samples to simulate challenging real-world environments. The Signal-to-Noise Ratio (SNR) for these noise injections is randomly sampled from a normal distribution with mean 10 dB and standard deviation 5 dB.

Furthermore, to improve the model's robustness to silence, we adopt a padding-before-noise strategy~\citep{an2025funaudio}. Specifically, for the 20\% training samples chosen for noise augmentation, we prepend and append short silence segments to the utterance prior to noise injection, where the duration of each silence segment is sampled from 0 to 1 second using a skewed $\mathrm{Beta}(1,3)$ distribution. This strategy helps mitigate hallucinations in both offline and streaming inference. It is particularly beneficial for streaming scenarios, where pauses between words or phrases may cause individual chunks to contain a non-negligible proportion of non-speech frames that can trigger erroneous outputs. By explicitly exposing the model to such cases during training, it learns to better distinguish speech from non-speech content, thereby reducing the risk of hallucinations.

\textbf{Training settings.}
The model is trained using the Adam optimizer~\citep{kingma2014adam} with cosine annealing and a 10k-step warm-up (except for RL). Our training corpus consists exclusively of Mandarin, Chinese dialects, English, and code-switched Mandarin–English speech data. Table~\ref{tab:training_detail} details the training data scale and maximum learning rate for each stage. 

\begin{table}[htbp]
    \centering
    \caption{Training details for all stages.}
    \label{tab:training_detail}
    
    \resizebox{0.8\linewidth}{!}{%
    \begin{tabular}{lcc}
        \toprule
        \textbf{Stage} & \textbf{Training Data} & \textbf{Maximum Learning Rate} \\
        \midrule
        Stage~1~Encoder pretraining & 560k hours & $5 \times 10^{-4}$ \\
        Stage~2~Alignment & 50k hours & $1 \times 10^{-3}$ \\
        Stage~3~IA-SFT & 560k hours & $1 \times 10^{-5}$ \\
        Stage~4~Late joint SFT & 560k hours & $1 \times 10^{-5}$ \\
        Stage~5~Context SFT & 50k hours & $1 \times 10^{-6}$ \\
        Stage~6~RL & 20k samples & $2 \times 10^{-6}$ \\
        \bottomrule
    \end{tabular}%
    }
\end{table}

\subsection{Inference}
\label{sec-Inference}


\subsubsection{Optimized Streaming Inference Pipeline}
\label{sec-Optimized_Streaming_Inference}

To achieve low-latency and high-throughput deployment in real-world streaming scenarios, NIM4-ASR adopts a decoupled inference architecture, allowing different modules to be deployed on separate accelerators to better utilize heterogeneous computing resources. The speech encoder is deployed on Triton Inference Server\footnote{\footnotesize \url{https://github.com/triton-inference-server/server}}, enabling dynamic batching across concurrent audio streams and significantly improving GPU utilization under high request concurrency. The adaptor and LLM decoder are served using a vLLM-based inference engine\footnote{\footnotesize \url{https://github.com/vllm-project/vllm}} that provides efficient KV-cache management. During inference, the encoder continuously processes incoming audio and transmits speech representations to the vLLM server, where they are projected into speech embeddings and appended to the LLM context. In addition, both the phoneme-level CTC head and the RAG module run on the CPU, where the CTC head produces phoneme hypotheses that are used for hotword retrieval.

To make the decoding pipeline more streaming-friendly, the prompt structure follows a fixed ordering. A static instruction prefix (e.g., system prompts and task instructions) is placed at the beginning of the context and can therefore be pre-computed and cached in the KV-cache before inference begins. Streaming speech embeddings are then appended incrementally as audio chunks arrive. Contextual information, including the instruction associated with the current decoding stage and hotwords retrieved by the RAG module, is appended after the speech embeddings. This ordering allows the static prefix to be cached once, while the speech embeddings are continuously prefetched during speech input. In the streaming output mode, intermediate decoding can be performed after each newly prefetched speech chunk. At the end of speech, all speech chunks have already been prefetched, and only the final instruction and accumulated hotword context need to be appended before final decoding, substantially reducing redundant KV-cache computation.

NIM4-ASR performs streaming recognition through incremental speech prefill and hypothesis refresh. The streaming encoder processes incoming speech with a chunk size of 640\,ms. Each chunk is encoded immediately, and the resulting speech representations are appended to the LLM context through streaming chunked prefill. During inference, the encoder caches representations from the previous 4 chunks, allowing the current chunk to attend to a limited left context while avoiding redundant computation over earlier audio segments. This design follows a cache-aware streaming strategy in which intermediate representations are reused across chunks rather than recomputed~\citep{noroozi2024stateful}. Meanwhile, the speech embeddings are incrementally accumulated in the LLM KV cache, enabling transcription hypotheses to be refreshed without re-prefilling the complete audio history.

For streaming output, the LLM updates the partial transcription after every 640\,ms speech chunk. Previously generated text is retained as a prefix, while the most recent 5 tokens can be rolled back and regenerated to correct unstable predictions near the speech chunk boundary. In addition, the hypotheses generated from the first two speech chunks are treated as unstable and can be revised more aggressively, mitigating errors caused by insufficient acoustic context at the beginning of an utterance. The phoneme-level CTC head and RAG module also operate continuously during streaming inference, with newly retrieved hotwords accumulated and deduplicated throughout the utterance. When the VAD module detects the end of speech, NIM4-ASR performs a second-pass final decoding to generate a stable final transcription. At this point, all speech embeddings have already been prefetched and retained in the LLM KV cache. Therefore, only the complete decoding instruction and the accumulated hotword context need to be appended before final decoding, allowing the second pass to start immediately with minimal additional prefill overhead. This design enables a low time-to-first-token (TTFT) through streaming hypothesis refresh while also reducing tail latency in second-pass final decoding, thereby ensuring both responsive streaming output and a stable final transcription. For offline inference, the complete audio is processed once by the encoder using the maximum chunk size, followed by a single-pass LLM decoding.

\subsubsection{Phoneme-based RAG for Hotword Customization}
\label{sec-hotword_RAG}

To enable efficient hotword customization, NIM4-ASR builds a phoneme-based hotword database with a corresponding retrieval algorithm, as illustrated in Figure~\ref{fig1}. Following prior work~\citep{an2025funaudio}, we preconvert each hotword text into a phoneme-token sequence and store it as a key-value pair, where the key is the phoneme sequence and the value is the corresponding hotword text. These phoneme sequences are first converted into discrete indices based on the phoneme vocabulary, and then restructured into a trie augmented with failure links using the Aho-Corasick automaton~\citep{aho1975efficient} algorithm. 
During streaming inference, the phoneme head attached to the encoder continuously generates phoneme hypotheses from newly arrived audio chunks. The hypotheses are converted into index sequences and scanned by the automaton incrementally, while newly retrieved hotwords are accumulated throughout the utterance and made available to subsequent decoding steps. For offline inference, the same retrieval procedure is applied once to the phoneme hypothesis of the complete utterance. When a partial match cannot be extended, the automaton follows the failure link to the longest valid suffix state instead of restarting the search from scratch, enabling all candidate hotwords to be retrieved with linear-time complexity in the hypothesis length.

To reduce redundant contextual hints, we apply a longest-match filtering strategy: shorter matches fully covered by longer spans are discarded, retaining only the longest entity. For example, if both the hotwords ``NIO'' and ``NIO House'' are matched in the same hypothesis, only ``NIO House'' is retained. The retrieved hotword texts are then concatenated and injected into the LLM prompt as contextual hints together with the speech embeddings, providing context-aware biasing for decoding. Owing to the storage efficiency of index-level mapping and the linear-time complexity of the Aho-Corasick automaton that depends only on query length rather than database size, the hotword database can easily scale to millions of entries while maintaining sub-millisecond retrieval latency per query.

It is worth noting that our hotword customization is designed to optimize the recognition of named entities such as location names and media titles, where the hotword database can be large and may contain numerous phonetically similar or even homophonous entries. To ensure retrieval precision under such large-scale settings, we adopt a hard-matching strategy in the RAG module, retrieving only exact phoneme-sequence matches rather than approximate ones or those with minimal edit distance. Empirically, retrieval misses are often less harmful than retrieval errors, since the LLM can still recover the correct entity from internal linguistic knowledge and context. By contrast, soft matching is more prone to introducing similar but incorrect hotwords, which can interfere with decoding even if the model is robust to noisy contextual hints to some extent.

\section{Evaluation}
\label{sec-Evaluation}

\subsection{Evaluation Setup}
\label{sec-Evaluation_Setup}

We evaluate NIM4-ASR on both public benchmarks and internal benchmarks to assess its performance across diverse domains.

\paragraph{Baseline Systems.}
We compare NIM4-ASR with several recent representative open-source LLM-based ASR models, including Fun-ASR-Nano~\citep{an2025funaudio}, GLM-ASR-Nano\footnote{\footnotesize \url{https://huggingface.co/zai-org/GLM-ASR-Nano-2512}}, Qwen3-ASR-1.7B~\citep{shi2026qwen3}, and FireRedASR2S-LLM~\citep{xu2026fireredasr2}. In addition, we also compare NIM4-ASR against large audio language models (LALMs) and multimodal LLMs with strong ASR capabilities, including Step-Audio2-Mini~\citep{wu2025step} and Qwen3-Omni-Instruct~\citep{xu2025qwen3}. While Fun-ASR, Qwen3-ASR and Qwen3-Omni support streaming inference, all baselines are evaluated in the offline setting for fair comparison. For NIM4-ASR, we report results under both offline and streaming inference settings.

\paragraph{Evaluation Metrics.}
We report Word Error Rate (WER) for English benchmarks, and Character Error Rate (CER) for Mandarin, Chinese dialect, lyrics, and code-switched Chinese-English benchmarks. As our internal benchmarks mainly consist of Mandarin speech, we use CER by default for internal evaluation. To minimize the influence of surface-level variation such as numeric expression formats and filler-word usage on evaluation statistics, we apply WeTextProcessing\footnote{\footnotesize \url{https://github.com/wenet-e2e/WeTextProcessing}}, a WFST-based toolkit for text normalization. This process may result in relatively lower absolute error rates across models, but it enables a fairer comparison of their intrinsic recognition capabilities. All baselines are reproduced following the official guidelines, and all transcriptions are normalized with the same pipeline to ensure consistent cross-system evaluation.

\paragraph{Public Benchmarks.}
Public evaluation datasets cover a wide range of speech recognition scenarios. English benchmarks include LibriSpeech~\citep{panayotov2015librispeech},  
VoxPopuli~\citep{wang2021voxpopuli}, 
and MLS-English~\citep{pratap2020mls}. 
Mandarin benchmarks include AISHELL-1~\citep{bu2017aishell}, AISHELL-2~\citep{du2018aishell}, AISHELL-2021-Eval\footnote{\footnotesize \url{https://aishelltech.com/aishell_2021_eval}}, WeNetSpeech~\citep{zhang2022wenetspeech}, and SpeechIO\footnote{\footnotesize \url{https://github.com/SpeechColab/Leaderboard}}. Chinese dialect evaluation includes WeNetSpeech-Chuan~\citep{dai2025wenetspeechchuan}, WeNetSpeech-Yue~\citep{li2025wenetspeechyue}, 
and KeSpeech~\citep{tang2021kespeech}. Additional challenging benchmarks include Mandarin-English code-switching speech from CS-Dialogue~\citep{zhou2025cs} and ASCEND~\citep{lovenia2022ascend}, as well as lyric transcription on M4Singer~\citep{m4singer}.

\paragraph{Internal Benchmarks.}
We further evaluate on a collection of internal benchmarks focused on realistic in-car spontaneous speech scenarios---a setting that differs markedly from conventional read-speech or conference-style corpora. These benchmarks mainly comprise instructional and conversational utterances that reflect real-world user interaction patterns, offering a more practical measure of ASR reliability in diverse in-car scenarios. All data were created by designing utterances grounded in real-world cockpit scenarios, and then collected through crowdsourced speaker recording.

\begin{itemize}[leftmargin=*]

\item \textbf{Point of Interest (POI)} data contains city-level POIs, which are derived from location names across different cities.

\item \textbf{Media} data involves media-related entities, including music titles, video titles, and radio program names.

\item \textbf{Device Control} data contains in-car control commands, such as vehicle setting adjustments and cockpit operation instructions. 

\item \textbf{Conversational} data includes two categories of conversational interactions: 
(1) Vehicle-domain chat data focuses on vehicle-related conversations such as in-car knowledge queries and assistant interactions; 
(2) Multi-domain chat data covers open-domain conversational queries across diverse domains including media, sports, healthcare, history, arts, literature, ecology, tourism, technology, science, culture, education, finance and entertainment.

\end{itemize}


\subsection{Evaluation Results}
\label{sec-Evaluation_Results}

\subsubsection{Public Benchmarks}
\label{sec-Public_Benchmarks}

Table~\ref{tab:main_benchmark} reports the comparison results on public benchmarks. For NIM4-ASR, we report both offline and streaming inference results. The offline setting reflects the upper-bound performance when full acoustic context is available, while the streaming setting evaluates real-time recognition.

Overall, NIM4-ASR shows strong competitiveness in the offline setting. It consistently outperforms baselines with smaller model sizes and achieves comparable or superior results against systems with more than 8B parameters. Across open-source benchmarks, NIM4-ASR delivers robust performance on Mandarin, dialectal speech, English, and code-switching. The main exception is meeting-style benchmarks, such as WeNetSpeech Meeting, where it performs slightly worse than competing models. This behavior is expected because NIM4-ASR is primarily optimized for streaming speech interaction scenarios that require low-latency responses to short and medium-length utterances. In contrast, long-form meeting transcription lies outside the primary design scope of the system and is correspondingly less represented in the training data.

Beyond the offline comparison, we find that NIM4-ASR also achieves satisfactory performance in the streaming mode, with only limited degradation relative to offline decoding. This can be attributed to two factors: first, the strict local alignment induced by CTC helps maintain stable acoustic representations under chunk-wise streaming inference; second, our dynamic chunk size and context length streaming training strategy enables the model to make robust predictions even with constrained acoustic context.

\begin{table*}[t]
\centering
\caption{
Comparison with recent advanced baselines on public benchmarks. All baseline systems are evaluated in offline mode. ``N/A'' denotes that a reliable result cannot be obtained under the official inference interface.
}
\label{tab:main_benchmark}

\resizebox{\textwidth}{!}{
\renewcommand{\arraystretch}{1.15}
\begin{tabular}{lcccccc|cc}
\toprule
  & \textbf{Fun-ASR} & \textbf{GLM-ASR} & \textbf{Qwen3-ASR} & \textbf{FireRedASR2S} & \textbf{Step-Audio2} & \textbf{Qwen3-Omni} & \multicolumn{2}{c}{\textbf{NIM4-ASR}} \\
  & \textbf{Nano} & \textbf{Nano} & \textbf{1.7B} & \textbf{LLM} & \textbf{Mini} & \textbf{Instruct} & \textbf{Offline} & \textbf{Stream} \\
  \midrule
  Model Size & 0.8B & 1.5B & 2.0B & 8B+ & 8B+ & 30B-A3B & 2.3B & 2.3B \\
  \midrule

\multicolumn{9}{l}{\textit{\textbf{Mandarin}}} \\

AISHELL-1 \textit{dev | test} & 1.59 | 1.81 & 2.40 | 2.41 & 1.40 | 1.51 & 0.60 | 0.64 & 0.76 | 0.81 & 0.86 | 0.92 & 0.43 | 0.57 & 0.43 | 0.60 \\
AISHELL-2-ios \textit{dev | test} & 2.62 | 2.73 & 3.21 | 3.45 & 2.41 | 2.60 & 2.07 | 2.08 & 2.24 | 2.29 & 2.11 | 2.31 & 2.28 | 2.43 & 2.33 | 2.49 \\
AISHELL-2021-Eval \textit{A | C | D}& 4.75 | 4.29 | 2.33 & 7.25 | 9.48 | 3.40 & 4.22 | 3.51 | 1.82 & 13.40 | 3.92 | 4.68 & 4.54 | 3.69 | 2.34& 5.19 | 3.34 | 1.66 & 3.12 | 1.51 | 1.81 & 3.28 | 1.63 | 2.22 \\
WeNetSpeech \textit{meeting | net} & 4.68 | 5.22 & 6.87 | 5.72  & 4.00 | 4.13 & 3.36 | 3.52 & 4.23 | 4.63 & 3.92 | 3.85 & 4.91 | 4.72 & 5.71 | 5.00 \\
SpeechIO  & 2.78 & 3.17  & 2.55 & 2.20 & 3.41  & 2.33 & 2.61 & 2.84 \\

\midrule
\multicolumn{9}{l}{\textit{\textbf{Chinese Dialect}}} \\

WeNetSpeech-Chuan \textit{easy | hard}  & 13.21 | 23.76 & 20.95 | 33.61 & 11.18 | 20.35 & 10.36 | 20.07 & 13.99 | 25.35 & 14.13 | 25.16 & 10.51 | 20.58 & 11.22 | 20.37 \\
WeNetSpeech-Yue \textit{short | long}  & 7.31 | 10.02 & 16.78 | 13.97 & 5.79 | 8.00 & 5.05 | 10.45 & 7.78 | 8.44 & 6.97 | 8.60  & 5.12 | 8.58 & 5.39 | 9.62 \\
KeSpeech  & 7.18 & 9.59  & 4.98 & 3.05 & 3.98 & 6.00 & 4.40 & 5.08 \\

\midrule
\multicolumn{9}{l}{\textit{\textbf{English}}} \\
LibriSpeech-dev \textit{clean | other} & 1.63 | 4.06 & 1.82 | 3.93 & 1.54 | 3.14 & 1.27 | 2.63 & 1.06 | 2.48 & 1.08 | 2.10 & 1.13 | 2.45 & 1.18 | 2.86  \\
LibriSpeech-test \textit{clean | other} & 1.63 | 4.35 & 1.96 | 4.29 & 1.56 | 3.49 & 1.29 | 2.97 & 1.22 | 2.61 & 1.15 | 2.38 & 1.19 | 2.53 & 1.29 | 2.92  \\
VoxPopuli \textit{dev | test} & 7.86 | 7.70 & 8.78 | 8.52 & 7.58 | 7.42 & 9.38 | 9.24 & 8.86 | 8.37 & 6.86 | 6.75 & 6.18 | 6.08 & 6.26 | 6.22  \\
MLS-English & 6.80 & 5.32  & 4.93 & 4.71 & 4.37 & 4.04 & 4.77 & 5.04  \\

\midrule
\multicolumn{9}{l}{\textit{\textbf{Mandarin-English Code-switch}}} \\
CS-Dialogue & 5.37 & 6.15  & 5.44 & 4.63 & 9.46 & 8.51  & 4.70 & 4.91 \\
ASCEND  & 11.91 & 12.29  & 10.87 & 10.22 & 13.50 & 18.68 & 11.46 & 11.85 \\

\midrule
\multicolumn{9}{l}{\textit{\textbf{Lyrics}}} \\
M4Singer  & 5.25 & 18.45  & 5.72 & N/A & 9.68 & 8.40 & 6.39 & 6.94 \\

\midrule\midrule
\multicolumn{9}{l}{\textbf{NIM4-ASR offline vs. Baselines}} \\
Win : Lose  & 23:2 & 25:0  & 18:7 & 12:12 & 17:8 & 14:11 &  - & - \\
\bottomrule

\end{tabular}
}
\end{table*}

\subsubsection{Internal Benchmarks}
\label{sec-Internal_Benchmarks}
  
\begin{table*}[ht]
\centering
\caption{Comparison with recent advanced baselines on internal benchmarks. All baseline systems are evaluated in offline mode. NIM4-ASR demonstrates consistent performance advantages on most internal benchmarks, as the evaluated content largely consists of long-tail named entities that open-source models rarely encounter during training.}

\label{tab:internal}

\resizebox{\textwidth}{!}{
\renewcommand{\arraystretch}{1.15}
\begin{tabular}{lcccccc|cc}
\toprule
 & \textbf{Fun-ASR} & \textbf{GLM-ASR} & \textbf{Qwen3-ASR} & \textbf{FireRedASR2S} & \textbf{Step-Audio2} & \textbf{Qwen3-Omni} & \multicolumn{2}{c}{\textbf{NIM4-ASR}} \\
 
 & \textbf{Nano}  & \textbf{Nano} & \textbf{1.7B} & \textbf{LLM} & \textbf{Mini} & \textbf{Instruct}  & \textbf{Offline} & \textbf{Stream} \\
\midrule
Model Size & 0.8B & 1.5B & 2.0B & 8B+ & 8B+ & 30B-A3B & 2.3B & 2.3B \\
\midrule

\multicolumn{9}{l}{\textit{\textbf{Point of Interest (POI)}}} \\
City A  & 7.07 & 14.68  & 9.14 & 8.54 & 9.41 & 9.67 & 3.86 & 3.85 \\
City B  & 8.50 & 15.75  & 10.59 & 10.43 & 11.67 & 11.73 & 4.86 & 4.94 \\
City C  & 7.60 & 17.55  & 10.01 & 10.17 & 11.35 & 12.18 & 3.77 & 3.81 \\
City D  & 7.42 & 17.91 & 9.77 & 9.51 & 11.55 & 10.86 & 4.10 & 4.17 \\

\midrule
\multicolumn{9}{l}{\textit{\textbf{Media}}} \\
Music  & 12.60 & 24.25 & 12.67 & 12.13 & 14.94 & 15.89 & 5.75 & 5.78 \\
Video  & 8.27 & 20.35 & 9.69 & 9.38 & 12.30 & 15.33 & 2.99 & 3.03 \\
Radio  & 13.69 & 19.82 & 10.51 & 11.84 & 14.21 & 17.91 & 1.21 & 1.17 \\

\midrule
\multicolumn{9}{l}{\textit{\textbf{Device Control}}} \\
Vehicle control & 4.74 & 8.78  & 5.31 & 4.52 & 4.97 & 4.18 & 1.88 & 1.78 \\

\midrule
\multicolumn{9}{l}{\textit{\textbf{Conversational}}} \\
Vehicle-domain chat \textit{easy | hard}  & 3.75 | 5.92 & 5.63 | 10.12 & 3.31 | 5.96 & 2.93 | 5.61 & 2.35 | 7.63 & 5.98 | 6.60 & 2.70 | 4.88 & 2.76 | 4.83 \\
Multi-domain chat  & 1.65 & 1.89  & 1.33 & 1.27 & 1.49 & 5.34 & 1.55 & 1.75 \\

\bottomrule
\end{tabular}
}
\end{table*}

Table~\ref{tab:internal} reports results on our internal benchmarks. Two benchmarks, POI and Media, are entity-intensive, comprising dense location names and media-related entities respectively. A key challenge in these domains is that many entities share similar or identical pronunciations, requiring the model to simultaneously resolve subtle acoustic differences and leverage contextual semantics to disambiguate competing candidates. NIM4-ASR achieves particularly strong performance on these benchmarks, driven primarily by comprehensive in-domain training data coverage, but also indicating that our training strategy effectively preserves both the encoder's fine-grained acoustic discriminability and the LLM's capacity for context-driven entity resolution.

Furthermore, NIM4-ASR also delivers clear improvements on both the vehicle control and vehicle-domain chat benchmarks. We attribute this gap primarily to the long-tailed nature of domain knowledge and terminology in general-purpose foundation models. By substantially increasing in-domain data coverage, NIM4-ASR achieves more reliable recognition of vehicle control commands and in-car assistant knowledge, thereby delivering a superior interaction experience within the vehicle cockpit. 
By contrast, on the multi-domain chat benchmark, spanning open-domain topics without vehicle-specific content, NIM4-ASR no longer leads but remains competitive. This demonstrates the model’s strong generalization ability: despite limited training data coverage in domains such as sports, healthcare, and finance, NIM4-ASR still maintains robust performance, indicating that its gains are not solely driven by domain-specific data expansion.



\subsubsection{Effectiveness of Hotword Customization}
\label{sec-Effectiveness_of_Hotword_Customization}

\begin{table}[H]
\centering
\caption{Effectiveness of phoneme-based hotword RAG on internal entity-intensive POI benchmarks. Recall here refers to the proportion of POI entities correctly recognized in the transcription output.}
\label{tab:hotword_effect}

\resizebox{0.51\columnwidth}{!}{
\begin{tabular}{lcc}
\toprule
 & 
   \textbf{Streaming} & \textbf{Streaming+RAG} \\
  \cmidrule(lr){2-2} \cmidrule(lr){3-3}
 & 
   CER / Recall & CER / Recall \\
\midrule

City A POI & 
   3.85 / 82.63 & 3.33 / 88.07 \\
City B POI & 
   4.94 / 77.48 & 4.31 / 83.60 \\


\bottomrule
\end{tabular}
}
\end{table}

Beyond its strong fundamental recognition capability, NIM4-ASR also provides an effective hotword customization mechanism. Through contextual hotword conditioning, NIM4-ASR can improve recognition accuracy for acoustically similar entity names, domain-specific terminology, and newly emerging expressions. To evaluate the effectiveness of the proposed hotword RAG mechanism, we focus on entity-intensive POI recognition scenarios, selecting benchmarks from two major cities and constructing city-specific retrieval databases, each comprising millions of location name–phoneme pairs. As shown in Table~\ref{tab:hotword_effect}, incorporating hotword context consistently improves streaming performance, demonstrating the effectiveness of our phoneme-based RAG retrieval mechanism and its practical benefit in entity-intensive recognition scenarios.

It is worth noting that, unlike previous work, we adopt exact matching rather than edit distance for retrieval. We argue that for the RAG module in LLM-based ASR systems, retrieval precision is more critical than recall, as the model's strong inherent recognition capability already serves as a reliable fallback when no hotword is retrieved. Moreover, pairing exact matching with the Aho-Corasick algorithm allows the hotword database to scale to millions of entries without additional retrieval overhead, avoiding the latency and precision degradation that typically follows vocabulary expansion.

\subsubsection{Effectiveness on Hallucination Mitigation}

\begin{table}[h]
  \centering
  \caption{Hallucination rate on different benchmark scenarios. “w/o RL” and “w/ RL” denote model after joint SFT and after the subsequent RL stage, respectively. For fair comparison, reported results for NIM4-ASR are obtained under offline inference.}
  \label{tab:hallucination_rate}
  \resizebox{0.85\columnwidth}{!}{%
  \begin{tabular}{lccccc}
  \toprule
  Model & Mandarin & Dialect & English & Code-switch & Lyrics\\
  \midrule
  Fun-ASR-nano      & 0.018\% & 0.217\% & 0.014\% & 0.397\% & 0.153\%\\
  GLM-ASR-nano      & 0.030\% & 0.201\% & 0.014\% & 0.315\% & 0.580\%\\
  Qwen3-ASR-1.7B    & 0.018\% & 0.120\% & 0.014\% & 0.345\% & 0.249\%\\
  FireRedASR2S-LLM  & 0.165\% & 0.298\% & 0.014\% & 0.335\% & 1.775\%  \\
  Step-Audio2-mini  & 0.020\% & 0.194\% & 0.014\% & 1.255\% & 0.390\%\\
  Qwen3-Omni-Inst   & 0.013\% & 0.370\% & \textbf{0.007\%} & 1.778\% & 0.129\%\\
  \midrule

  NIM4-ASR (offline w/o RL) & 0.003\% & 0.122\% & \textbf{0.007\%} & \textbf{0.261\%} & 0.215\%\\
  NIM4-ASR (offline w/ RL) & \textbf{0.002\%} & \textbf{0.117\%} & \textbf{0.007\%} & \textbf{0.261\%} & \textbf{0.081\%} \\
  \bottomrule
  \end{tabular}%
}
  
\end{table}

Beyond recognition performance, NIM4-ASR demonstrates strong hallucination suppression. We compare all baseline models and NIM4-ASR in terms of hallucination rate across five distinct scenarios, where the rate for each scenario is defined as the ratio of hallucinated samples to total samples aggregated over all benchmarks within that scenario. Specifically, a sample is classified as hallucinated if its transcription exceeds the ground-truth length by over 50\% with negligible lexical overlap. Notably, we exclude three benchmarks: WeNetSpeech Meeting, SpeechIO, and MLS-English from this evaluation, as no hallucinated samples are observed across any model; we additionally exclude WeNetSpeech Net, as its prevalence of unreliably annotated short samples inflates hallucination rates across all models.

As shown in Table~\ref{tab:hallucination_rate}, NIM4-ASR achieves substantially lower hallucination rates compared to all baseline models. Attributed to our training paradigm design and noise data augmentation, the model after joint SFT already exhibits a low hallucination rate: only marginally above the best-performing baseline on Dialect and Lyrics benchmarks. After the RL stage, the hallucination rate is further reduced, achieving the lowest average across all five scenarios; most notably, on Mandarin benchmarks, NIM4-ASR attains a hallucination rate of 0.002\%, substantially below all baselines.

\subsubsection{Effectiveness of RL}
\label{sec-Effectiveness_of_RL}

\begin{table}[h]
\centering
\caption{Effectiveness of the RL stage under different inference settings. “w/o RL” and “w/ RL” correspond to the model after joint SFT and after the RL stage, respectively.}
\label{tab:ablation_rl}

\renewcommand{\arraystretch}{1}
\resizebox{1.0\columnwidth}{!}{
\begin{tabular}{lcccccc} 
\toprule

& \multicolumn{2}{c}{\textbf{w/o RL}} 
& \multicolumn{2}{c}{\textbf{w/ RL}} 
& \multicolumn{2}{c}{\textbf{RL Gain (+/-)}}\\
\cmidrule(lr){2-3} \cmidrule(lr){4-5} \cmidrule(lr){6-7} 

& \textbf{Offline} 
& \textbf{Streaming}
& \textbf{Offline} 
& \textbf{Streaming}
& \textbf{Offline} 
& \textbf{Streaming} \\ 

\midrule
Open-source Mandarin Avg.& 2.71 & 2.80 & 2.44 & 2.65 &-0.27&-0.15 \\
Open-source English Avg.& 3.55 & 3.76 & 3.48 & 3.68 &-0.07&-0.08  \\
Mandarin–English Code-switch Avg.&8.39&8.62&8.08&8.38&-0.31&-0.24 \\
Internal Mandarin Avg. & 3.57 & 3.69 & 3.41 &3.44& -0.16&-0.25 \\

\bottomrule
\end{tabular}
}
\end{table}

We further ablate the RL stage to assess its contribution. As shown in Table~\ref{tab:ablation_rl}, incorporating RL yields consistent improvements under both offline and streaming settings, with the most substantial gains observed on Mandarin and code-switching benchmarks. A key contributing factor is the high prevalence of homophone and near-homophone confusions in these scenarios, which token-level teacher-forcing does not directly penalize effectively. By contrast, RL optimizes sequence-level rewards over complete decoding trajectories, explicitly penalizing sentence-level error propagation induced by phonetic confusion, reinforcing entity and phrase-level consistency, and mitigating exposure bias~\citep{chen2025reinforcement}. In code-switching scenarios, acoustic ambiguity at language-switch boundaries and cross-lingual entity competition are particularly pronounced; sequence-level feedback from RL can more effectively suppress erroneous language continuation and improve overall transcription consistency under mixed Mandarin–English conditions.

\section{Conclusion}
\label{sec-Conclusion}

In this work, we revisit LLM-based ASR from a deployment-oriented perspective and identify three obstacles that continue to hinder practical adoption: limited downward scalability arising from cross-modal alignment overhead, hallucination induced by representation drift during joint optimization, and the lack of production-ready mechanisms for hotword customization. NIM4-ASR addresses these challenges through targeted architectural design and a multi-stage training paradigm. By explicitly anchoring each training stage to the functional boundaries of its constituent modules, NIM4-ASR improves parameter utilization efficiency and mitigates hallucinations under acoustically ambiguous conditions, thus building a more stable foundation for LLM-based streaming speech recognition.
Building on this principle, NIM4-ASR further incorporates a real-time streaming inference pipeline and phoneme-level RAG to enable million-scale hotword customization. Extensive evaluation on 25 benchmarks demonstrates that NIM4-ASR achieves SOTA performance on several benchmarks with only 2.3B parameters, while maintaining low-latency streaming capability and clear advantages in entity-intensive scenarios. Overall, these results suggest that advancing LLM-based ASR relies not only on scaling model capacity, but more importantly on co-designing model architecture, training objectives, and inference strategies. NIM4-ASR thus provides a practical solution for building efficient, robust, and customizable LLM-based ASR systems for real-time speech interaction.

\section{Limitations and Future Work}
\label{sec-Limitations_and_Future_Work}

Although NIM4-ASR has demonstrated strong recognition performance and practical effectiveness, several key issues remain to be addressed in the next stage of system iteration. First, the current model supports only Mandarin, English, and a limited set of Chinese dialects, leaving broader multilingual and dialectal coverage as an important direction for future work. Second, the current model uses only retrieved hotwords as contextual input and does not yet incorporate conversation history, leaving room for improvement in cross-turn transcription consistency in multi-turn interaction scenarios. In addition, the gains brought by RL are not yet sufficiently stable, suggesting that further optimization is needed in both algorithm design and reward formulation.
In future work, we plan to focus on the following directions:

\begin{itemize}[leftmargin=*,itemsep=2pt]
\item (1) Expanding support for more languages and Chinese dialects, and developing more adaptive hotword customization mechanisms for dialectal and accented speech.

\item (2) Incorporating conversation history as additional contextual information to improve cross-turn transcription consistency in multi-turn interaction scenarios.

\item (3) Further improving streaming inference efficiency and enabling scalable RAG acceleration under high-concurrency deployment settings.

\item (4) Refining the RL algorithm and reward design to further improve system robustness and reduce hallucinations.
\end{itemize}

\bibliographystyle{plainnat}  

\bibliography{references}  

\end{document}